\documentclass[aps,pre,twocolumn,superscriptaddress,showpacs,showkeys,floatfix]{revtex4}%

\usepackage{epsfig}
\usepackage{times}
\usepackage{amssymb}
\usepackage{colordvi}
\usepackage{color}     
\usepackage{amsbsy}
\usepackage{amsmath}
\usepackage{amsbsy} 	
\usepackage{latexsym}
\usepackage{graphicx}

\renewcommand{\pdfliteral}[1]{}

\newcommand{\blackcolor}{\pdfliteral{1 1 1 1 k}}
\newcommand{\DarkBlue}[1]{{\pdfliteral{0.98 0.13 0    0.63 k}#1}\blackcolor}   
\renewcommand{\Blue}[1]{{\pdfliteral{0.98 0.13 0    0.10 k}#1}\blackcolor}

  \newcommand{\commentcolor}{\pdfliteral{0.0 0.83 0 0.10 k}}

  \newcommand{\hidecolor}{\pdfliteral{0.3 0.3 0.3 0 k}}
  \newcommand{\addcolor}{\protect\pdfliteral{0.98 0.13 0 0.63 k}}

\newcommand{\RESTORECOLOR}{}   

\newcommand{\FINALREF}[1]{\DarkBlue{\ref{#1}}\RESTORECOLOR}

{\rm \addcolor $\bullet|$ \blackcolor}                

\newenvironment{comment}[1]
{\commentcolor$|\bullet$\commentcolor \em [#1]}  
{\rm \commentcolor $\bullet|$ \blackcolor}		

\newcommand{\HIDDEN}[1]{}
                          {]\end{quotation}\blackcolor\renewcommand{\RESTORECOLOR}{}\rm} 

\newcommand{\stl}[1]{{\hspace{0.2em}
\stackrel{ \setbox0=\hbox{\hspace{0.06em}$\displaystyle
#1$\hspace{0.06em}} \setbox1=\hbox{\vrule width\wd0 height0.08ex depth0pt} \vrule width0.08ex
height0.08ex depth0.475ex \box1 \vrule width0.08ex height0.08ex
depth0.475ex } {#1}\hspace{0.2em}}}

 \newcommand{\X}{{\bf X}}
 \newcommand{\ave}[1]{\left\langle #1\right\rangle}
 \newcommand{\avee}[1]{\langle #1\rangle}
 
 \newcommand{\bea}{\begin{eqnarray}}
 \newcommand{\eea}{\end{eqnarray}}
 \newcommand{\be}{\begin{equation}}
 \newcommand{\ee}{\end{equation}}

 \newcommand{\dphi}{\delta\varphi}
 \newcommand{\dtmicro}{\partial_{t'}^{\rm mic}}
 \newcommand{\dtmacro}{\partial_{t'}^{\rm mac}}
 \renewcommand{\k}{{\bf k}}
 
 \newcommand{\mbf}[1]{\mbox{\boldmath{$#1$}}}
 
 \newcommand{\q}{\tilde{\bf q}}
 \newcommand{\x}{{\bf x}}
 \newcommand{\ONE}{\mathbf{I}}

 \newcommand{\e}{{\bf e}_\parallel}

 \newcommand{\gammapara}{c_\parallel}

 \newcommand{\beas}[1]{\begin{subequations}\label{#1}\bea}
 \newcommand{\eeas}{\eea\end{subequations}}
 
 \newcommand{\gammaperpnew}{c_\perp}

 \newcommand{\df}{\delta f}

 \renewcommand{\c}{{\bf c}}
 \newcommand{\cc}{c}
 \newcommand{\kB}{k_{\rm B}}
 \renewcommand{\q}{{\bf q}}

 \newcommand{\n}{{n}}
 \newcommand{\T}{{T}}
 \renewcommand{\u}{{\bf u}}
 
 \renewcommand{\dtmicro}{\partial_{t}}
 \renewcommand{\dtmacro}{\partial_{t}}
 \newcommand{\E}{{\bf e}_\perp}
 \newcommand{\Ephi}{{\bf e}_\phi}

 \renewcommand{\aa}{\mbf{\omega}}
 \newcommand{\aaa}{\omega}
 \newcommand{\gammao}{c_\phi}
 \newcommand{\xfull}{\x_{\rm large}}
 \newcommand{\xlarge}{\xfull}
 
 \newcommand{\aveeGMS}[1]{\avee{#1}}
 
 \renewcommand{\Re}{{\rm Re}}
 \renewcommand{\Im}{{\rm Im}}
 
 \newcommand{\aveNOGM}[1]{\ave{#1}}

 \renewcommand{\parallel}{{\mbox{\tiny$\|$}}}
 \renewcommand{\perp}{{\mbox{\tiny$\bot$}}}

 \newcommand{\stress}{\mbf{\sigma}}

 \newcommand{\pard}[2]{\frac{\partial #1}{\partial #2}}
 
 \newcommand{\showfootnote}[1]{#1}
 
 \newcommand{\plus}{\oplus}
 \newcommand{\minus}{\ominus}
 \newcommand{\plusminus}{\otimes}

\begin{document}

\title{Exact linear hydrodynamics from the Boltzmann equation
}
\date{ {\small 2008-01-18 09:47:09} 
} 

\author{I.V. Karlin} \email{il.karlin@lav.mavt.ethz.ch} 
\affiliation{Aerothermochemistry and Combustion Systems Lab, ETH Z\"urich, CH-8092 Z\"urich, Switzerland and\\School of Engineering Sciences, University of Southampton, SO17 1BJ, Southampton, United Kingdom}

\author{M. Colangeli}  
\affiliation{Polymer Physics, Department of Materials and Materials Research Center, ETH Z\"urich, CH-8093 Z\"urich, Switzerland}

\author{M. Kr\"oger}  \homepage{www.complexfluids.ethz.ch}
\affiliation{Polymer Physics, Department of Materials and Materials Research Center, ETH Z\"urich, CH-8093 Z\"urich, Switzerland}


\newcommand{\INLINEFIGURE}[1]{#1}
\begin{abstract}
 	%
 Exact (to all orders in Knudsen number) equations of linear hydrodynamics are derived from the Boltzmann kinetic equation with the Bhatnagar-Gross-Krook collision integral. 
 	%
 	%
 	%
 	%
 The exact hydrodynamic equations are cast in a form 
 which allows us to immediately prove their
 hyperbolicity, stability, and existence of an $H$-theorem.

\end{abstract}


\maketitle


 
 Hydrodynamics assumes that a state of a fluid is solely described
 by five fields: density, momentum and temperature. Derivation of
 the Navier-Stokes-Fourier (NSF) hydrodynamic equations from the
 Boltzmann kinetic equation as the first-order approximation in
 the Knudsen number (ratio between mean free path and a flow
 scale) by Enskog and Chapman is a textbook example of a success of
 statistical physics \cite{chapman}. 
 Recent renewed interest to the problems beyond the standard hydrodynamics is due, in particular, to flow simulation and experiments at a micro- and nano-scale \cite{Karniadakis,Ansumali07,yakhot07}.
 However, almost a century of
 effort to extend the hydrodynamic description beyond the NSF
 approximation failed 
 even in the case of small deviations around
 the equilibrium.
 	%
 	%
 	%
 In order to appreciate the problem, let us remind that, in the NSF
 approximations, the decay rate of the hydrodynamic modes is
 quadratic in the wave vector, ${\rm Re}(\omega)\sim -k^2$, and is unbounded.
 On the other hand, Boltzmann's collision term
 features equilibration with finite characteristic rates.
 This ``finite collision frequency" is obviously incompatible with
 the arbitrary decay rates in the NSF approximation:
 intuitively, hydrodynamic modes at large $k$ cannot relax
 faster than the collision frequency. Now, the
 classical method of Enskog and Chapman extends the hydrodynamics
 beyond the NSF in such a way that the decay rate of the next order
 approximations (Burnett and super-Burnett) are polynomials of 
 higher order in $k$. In such an extension, relaxation rate may
 become completely unphysical (amplification instead of
 attenuation), as first shown by Bobylev \cite{Bobylev82} for a
 particular case of Maxwell molecules. This indicates inability of
 the Chapman-Enskog method to tackle the above problem, and
 non-perturbative approaches are sought. 
 The problem of exact hydrodynamics has been studied in depth recently for toy (finite-dimensional) models - moment systems of Grad - in \cite{Karlin96,matteo1,matteo2}, and many remarkable results were obtained. In particular, in \cite{matteo1,matteo2} it was shown that the exact hydrodynamic equations are hyperbolic and stable for all wave numbers. However, 
 for ``true" kinetic equations such questions remain open.
 
          \INLINEFIGURE{
          \begin{figure}[tbh]
          \begin{center}
          \includegraphics[width=8.5cm, angle=0]{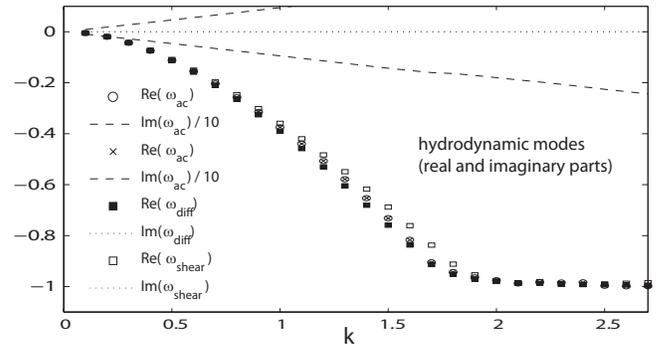}
          \caption{{\footnotesize Exact hydrodynamic modes $\omega$ of the Boltzmann-BGK kinetic equation as a function of wave number $k$ (two complex-conjugated acoustic modes $\omega_{\rm ac}$, twice
degenerated shear mode $\omega_{\rm shear}$ and thermal diffusion mode $\omega_{\rm diff}$). The non-positive decay rates $\Re(\omega)$ attain the limit of collision frequency ($-1$) as $k\to\infty$.
}}
          \label{2007jan18_Fig1}
          \end{center}
          \end{figure}
          }

 In this Letter we derive  exact hydrodynamic equations from
 the linearized Boltzmann equation with the Bhatnagar-Gross-Krook
 (BGK) collision term. This kinetic equation remains popular in applications \cite{Cercignani}, and features a single relaxation rate. The result for the hydrodynamic modes is
 demonstrated in Fig.\ \FINALREF{2007jan18_Fig1}. It is clear
 from Fig.\ \FINALREF{2007jan18_Fig1} that the relaxation of none of the hydrodynamic modes is faster than $\omega=-1$ which is the collision frequency in
 the units adopted in this paper. Thus, the result for the exact
 hydrodynamics indeed corresponds to the above intuitive picture. Below,
 we apply the method of invariant manifold \cite{Gorban94} to derive the hydrodynamic equations. 
 The non-perturbative derivation is made possible with an optimal
 combination of analytical and numerical approaches to solve the
 invariance equation.
 
 Point of departure is the linearized Boltzmann-BGK equation for
 the deviation $\triangle f = f-f^{\rm GM}$ of the distribution
 function $f$ from a global Maxwellian 
 $f^{\rm GM}(\cc^2)=\pi^{-3/2}e^{-\cc^2}$. In the reciprocal space, it reads,
 	%
 \be
  \dtmicro \triangle f = -i\k\cdot\c \,\triangle f - \df; \qquad \df = f - f^{\rm LM}, \label{boltz}
 \ee
 with the wave vector $\k=k\,\e$ defining $\e$, $k\equiv|\k|$, peculiar velocity $\c$ 
 and time $t$. All quantities are considered dimensionless, 
 i.e., reduced with the units of the relaxation time $\tau$,
 the thermal velocity $\sqrt{2\kB T/m}$ and mass $m$ of the particle.
 In (\FINALREF{boltz}), the linearized local Maxwellian 
 is  $f^{\rm LM}=f^{\rm GM}(1+\varphi_0)$ 
 where
 $1+\varphi_0 = \avee{1}_f + 2\c\cdot\avee{\c}_f + \frac{2}{3}(c^2-\frac{3}{2}) \avee{c^2-\frac{3}{2}}_f$.
 Averages are defined for arbitrary $\aa$ via $\avee{\aa}_f \equiv \int \aa f\,d^3\cc$, and
 we introduce pertinent quantities which characterize deviation from the global equilibrium:
 $\n \equiv \avee{1}_f-1$ (density perturbation), $\u\equiv \avee{\c}_f$ (velocity perturbation)
 and $\T \equiv \frac{2}{3}\avee{c^2-\frac{3}{2}}_f$ (temperature perturbation).
 Since the scalar product between $\k$ and $\c$ appears
 in (\FINALREF{boltz}), the distribution function offers symmetry with respect to the $\e$-axis,
 which is not uniaxial in case $\u$ is not collinear with $\e$. We denote 
 the two components of the mean velocity as $u^\parallel = \u\cdot\e$ and 
 $u^\perp=\u\cdot\E$,
 where the unit vector $\E$ belongs to the intersection of the plane perpendicular to $\k$ and the plane spanned by $\k$ and $\u$,
 so that $\u=u^\parallel\e+u^\perp\E$.
 	%
 Equations of change for  moments $\avee{\aa}$ are obtained 
 by integration of the weighted (\FINALREF{boltz}) over $d^3\cc$
 as
 \be
  \dtmicro \ave{\aa}_{\triangle f} = -i\k\cdot\ave{\c\,\aa}_{\triangle f} - \ave{\aa}_{\df}.
 \label{eocmom}
 \ee
 	%
 	%
 In order to calculate such averages, we can switch to spherical coordinates.
 For each (at present arbitrary) wave vector, we choose the coordinate system in such a way
 that its $z$-direction aligns with $\e$. We can then
 express $\c$ in terms of its norm $\cc$, a vertical variable $z$
 and plane vector $\Ephi$ 
 (azimuthal angle $\Ephi\cdot\E=\cos\phi$)
 for the present purpose as
 $\c/\cc = \sqrt{1-z^2}\,\Ephi + z \,\e$.
 We could have equally chosen a fixed coordinate system in the plane orthogonal to $\k$, and two fields 
 instead of $u^\perp$ plus an angle, viz. $u_x^\perp+iu_y^\perp=e^{i\phi}u^\perp$. 
 Due to isotropy, 
 $u^\perp$ alone fully represents the twice degenerated (shear) dynamics.
 In order to simplify notation and compute the dynamics of all five fields
 we introduce a four-dimensional vector of hydrodynamic fields,
 $\x\equiv (x_1,x_2,x_3,x_4) = (\x^\parallel,x_4)$ with $\x^\parallel\equiv(\n,u^\parallel,\T)$ and $x_4=u^\perp$. 
 Then $\varphi_0$ 
 takes a simple form, $\varphi_0 = \X^0\cdot\x$. 
 The vector $\X^0$ can immediately
 be read off, we have $\X^0(\cc,z)=(1,2\gammapara,\cc^2-\frac{3}{2},2\gammao)$, where we introduced,
 for later use, the abbreviations
 \be
  \gammapara \equiv \c\cdot\e
 ,\quad \gammao \equiv \c\cdot\E,
  \quad \gammaperpnew \equiv \frac{\gammao}{\E\cdot\Ephi}, 
 \label{defc}
 \ee
 such that $i\k\cdot\c=ik\gammapara$, $\c=\gammaperpnew\Ephi + \gammapara \e$
 with $\gammapara=cz$ and $\gammaperpnew=c\sqrt{1-z^2}$, contrasted by
 $\gammao$ (and $\Ephi$), do not depend on the azimuthal angle.
 Similarly, we introduce yet unknown fields $\delta \X(\c,\k)$ which characterize the
 nonequilibrium part of the distribution function, 
 $\dphi = \df/f^{\rm GM}$
 in terms of the hydrodynamic fields $\x$ themselves,
 \be
 \label{expanded}
 \delta \varphi=\delta \X\cdot \x = \delta X_1\n +\delta X_2u^\parallel 
 + \delta X_3\T + \delta X_4 u^\perp,
 \ee
  where 
 $\delta X_4$ factorizes as $\delta X_4(c,z,\phi) = 2\delta Y_4(c,z) \,\Ephi\cdot\E$.
 This ``eigen''-closure (\FINALREF{expanded}) which formally and very generally addresses the 
 fact, that we wish to {\em not} include other than hydrodynamic variables 
 implies a closure between moments of
 the distribution function, to be worked out in detail below. 
 It assumes the existence of an invariant 
 manifold, and the hydrodynamic fields as slow variables which leave the higher moments
 ``slaved''. 
 In order to ensure these
 contributions to not interfere with the local Maxwellian, one has the freedom to require 
 $\avee{\X^0}_{\df}=0$ without producing any limitation, i.e., by keeping
 $\x$ to be defined through the local Maxwellian part of the distribution function. 
 Using the above form for $\delta f$ in (\FINALREF{boltz}), and
 using the canonical abbreviation $\triangle \X \equiv \X^0 + \delta \X$, 
 yields  
 \be
  \triangle \X \cdot \dtmicro \x
 = -ik\,\gammapara\,\triangle \X \cdot\x - \delta \X\cdot\x,
  \label{macf}
 \ee
 which is a nonlinear integral equation for the unknown fields $\delta\X$, because 
 $\dtmicro \x$ has to be replaced by the right hand side of (\FINALREF{eocmom}), 
 for a suitably chosen vector $\aa$ fulfilling 
 $\avee{\aa}_{\triangle f}=\x$. 
 Here, $\aa$ is similar with $\X^0$ and differs
 from $\X^0$ mainly because of conventions for prefactors in the temperature and velocity definitions,
 $\aa = (1,\gammapara,\frac{2}{3}(\cc^2-\frac{3}{2}),\gammao)$.  
 Within the same eigen-closure, Eq.~(\FINALREF{eocmom}) is linear in $\x$ 
 and hence written as
 \be
  \dtmacro \x = {\bf M}\cdot\x.
 \label{dtx}
 \ee

 \begin{table}
  \begin{tabular}{c@{\quad}c@{\quad}c@{\quad}c}
  \hline\hline
 $\sigma^\parallel_1$ &  $\sigma^\parallel_2$ &$\sigma^\parallel_3$
 &$\sigma_4$\\
 $\aveeGMS{\lambda^\parallel \delta X_1}$ &
 $\aveeGMS{\lambda^\parallel  \delta X_2}$ &
 $\aveeGMS{\lambda^\parallel \delta X_3}$ &
 $\aveeGMS{ \gammapara\gammaperpnew \delta Y_4}$ \\
 $-k^2 B$ & $ik A$ & $-k^2 C$ & $ik D$ \\
 $B_0=-\frac{1}{3}$ & $A_0=-\frac{2}{3}$ & $C_0=0$ &
 $D_0=-\frac{1}{2}$ \\
 real, $\plus$ & imag,$\plus$ & real,$\plus$ & imag,$\minus$ \\
  \hline\hline
 $q^\parallel_1$ &$q^\parallel_2$ &$q^\parallel_3$ &$q_4$ \\
 $\aveeGMS{\gamma^\parallel\delta X_1}$ &
 $\aveeGMS{ \gamma^\parallel\delta X_2}$ &
 $\aveeGMS{\gamma^\parallel \gammapara  \delta X_3}$ &
 $\aveeGMS{(\cc^2-\frac{5}{2}) \gammaperpnew\delta Y_4}$ \\
 $ikX$ & $-k^2 Z$ & $ikY$ & $-k^2 U$ \\
 $X_0=0$ & $Z_0=-\frac{1}{6}$ & $Y_0=-\frac{5}{4}$ &
 $U_0=\frac{1}{2}$
 \\
 imag,$\minus$ & real,$\minus$ & imag,$\minus$ & real,$\plus$\\
 \hline\hline
  \end{tabular}
 	  \caption{
 Symmetry adapted components of (nonequilibrium) stress tensor $\stress$ and heat flux $\q$, 
 introduced in (\FINALREF{stress}) and (\FINALREF{heat}), respectively.
 Row 2: Microscopic expression of these components (averaging with the global Maxwellian).
        Short-hand notation used: 
        $\lambda^\parallel=\gammapara^2-\frac{\cc^2}{3}$ and 
        $\gamma^\parallel= (\cc^2-\frac{5}{2}) \gammapara $.
 Row 3: Expression of the components in terms of (as we show, real-valued) functions $A$--$Z$ (see text).
 Row 4: Values of functions $A$--$Z$ at $k=0$. 
        These values recover hydrodynamic equations up to Burnett approximation.
 Row 5: Parity with respect to $z$ -- symmetric ($\plus$) or antisymmetric ($\minus$) -- 
        of the part of the corresponding $\delta X$ entering the averaging in row 2, 
        and whether this part is imaginary or real-valued (see Fig.~\FINALREF{2007jan18_Fig2}). Row 3 is an immediate
 consequence of row 5.
 }
 	  \label{tab1}
 	  \end{table}
 
 

 The  matrix ${\bf M}$ solely depends on the non-hydrodynamic fields, the
 heat flux $\q \equiv \avee{\c(\cc^2-\frac{5}{2})}_{f}$
 and the stress tensor $\mbf{\sigma}\equiv \avee{\stl{\c\c}}_{f}$, where 
 $\stl{{\bf s}}$ denotes the symmetric traceless part of a tensor ${\bf s}$
 \cite{matteo2,mkbook},
 $\stl{{\bf s}}=\frac{1}{2}({\bf s}+{\bf s}^T) - \frac{1}{3}{\rm tr}({\bf s})\ONE$. 
 Using (\FINALREF{expanded}), the stress tensor and heat flux uniquely
 decompose as follows
 \beas{sq}
  \mbf{\sigma} &=& \sigma^\parallel\,\frac{3}{2}\stl{\e\e} + \sigma^\perp\,2\stl{\e\E}, \label{stress} \\
  \q &=& q^\parallel\,\e + q^\perp\,\E \label{heat}, 
 \eeas with the moments
 $\sigma^\parallel = (\sigma^\parallel_1,\sigma^\parallel_2,\sigma^\parallel_3)\cdot\x^\parallel$ 
 and $\sigma^\perp = \sigma_4x_4$, and similarly for $\q$ (see Row 2 of Tab.~\FINALREF{tab1}). 
 The prefactors arise from the identities $\stl{\e\e}:\e\e=\frac{2}{3}$ and $\stl{\e\e}:\e\E=\frac{1}{2}$.
 The appearance of $\delta Y_4$ rather than $\delta X_4$ in the expression for the
 orthogonal moment (in Tab.~\FINALREF{tab1}) reflects the fact that we 
 have already integrated out  
 the angular variable, $\int_0^{2\pi} \Ephi\Ephi\cdot\E\,d\phi = \pi\E$. 
 	%
 We note in passing that, while the stress tensor has, in general, three different eigenvalues, 
 in the present symmetry adapted  
 coordinate system it exhibits a vanishing first normal stress difference.
 	%
 Since the integral kernels of all moments in (\FINALREF{sq}) do not depend on the azimuthal angle, these
 are actually two-dimensional integrals over $c\in[0,\infty]$ and $z\in[-1,1]$, weighted by
 $2\pi \cc^2 f^{\rm GM}(\cc^2)\delta \X_\mu$.

 Stress tensor and heat flux can yet be written
 in an alternative form which is defined by Row 3 of Tab.~\FINALREF{tab1}. As we will see later on, 
 due to basic symmetry considerations, 
 the hereby introduced functions $A$--$Z$ 
 are real-valued. 
 	%
 We postpone the related proof, 
 and proceed by using these functions $A$--$Z$ 
 to split ${\bf M}$ into parts as
 ${\bf M} = \Re({\bf M}) - i \,\Im({\bf M})$ with 
 \bea
  \Re({\bf M}) &=& k^2 \left(\begin{array}{cccc}
  0 & 0 & 0 & 0 \\
 0 & A & 0 & 0 \\
 \frac{2}{3} X & 0 & \frac{2}{3}Y & 0 \\
 0 & 0 & 0 & D
 \end{array}\right), \label{checkerboard} \\
  \Im({\bf M}) &=& k \left(\begin{array}{cccc}
  0 & 1 & 0 & 0 \\
 \frac{1}{2}\!-\!k^2 B & 0 & \frac{1}{2}\!-\!k^2 C & 0 \\
 0 & \frac{2}{3}(1\!-\!k^2 Z) & 0 & 0 \\
 0 & 0 & 0 & 0
 \end{array}\right). \nonumber
 \eea
 
          \INLINEFIGURE{
          \begin{figure}[tbh]
          \begin{center}
          \includegraphics[width=8.5cm, angle=0]{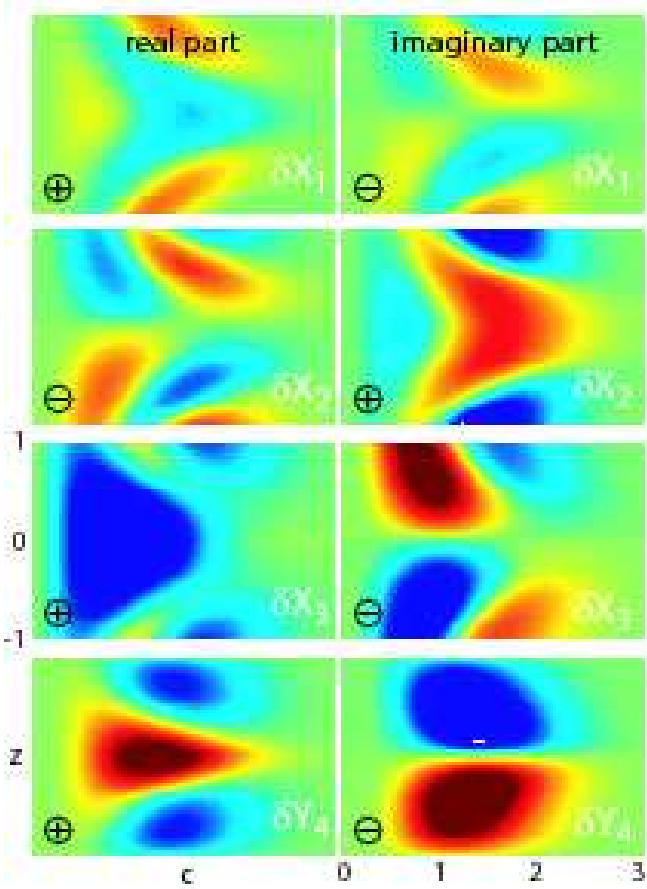}
          \caption{{\footnotesize (Color online) Sample distribution function $f(\c,\k)$ at $k=1$, fully characterized by the four quantities $\delta X_{1,2,3}(c,z)$ and $\delta Y_4(c,z)$. Shown here are both their real (left) and imaginary parts (right column). In order to improve contrast, we actually plot $\ln |1+f^{\rm GM}\delta X_\mu|$ multiplied by the sign of $\delta X_\mu$. Same color code for all plots, ranging from $-0.2$ (red) to $+0.2$ (blue). 
}}
          \label{2007jan18_Fig2}
          \end{center}
          \end{figure}
          }
 
 Note that the checkerboard structure of the matrix ${\bf M}$ (\FINALREF{checkerboard}) 
 is particularly useful for studying properties of the hydrodynamic equations (\FINALREF{dtx}),
 such as hyperbolicity and stability (see \cite{matteo1,matteo2} and below),
 once the functions $A$--$Z$ are explicitly evaluated. For that, we still require $\delta \X$. 
 	%
 	%
 Combining (\FINALREF{macf}) and (\FINALREF{dtx}), and requiring that the result holds for any $\x$ (invariance condition), we obtain a closed, singular integral equation (invariance equation) for complex-valued $\delta \X$,
 \be
   \delta \X = \X^0\cdot\left( {\bf M}+ [i k \gammapara + 1 ]\ONE\right)^{-1}-\X^0.
 \label{eqsolve}
 \ee
 Notice that $\delta \X$ vanishes for $k=0$, and that (\FINALREF{eqsolve}) 
 is supplemented with the basic constraint $\avee{\X^0}_{\df} = 0$, 
 or equally, vanishing Lagrange multipliers (matrix) $\avee{\X^0\,\delta \X}$, which, however,
 is automatically dealt with if we only evaluate anisotropic (irreducible) moments with $\df$, such 
 as those listed in Tab.\ \FINALREF{tab1}. 
 The implicit equation (\FINALREF{eqsolve}) 
 is identical with the eigen-closure (\FINALREF{expanded}), and is our main and practically useful 
 result, with ${\bf M}$ from (\FINALREF{checkerboard}), $\gammapara$, $\X^0$, and $A$--$Z$ defined in 
 and just before (\FINALREF{defc}) and Tab.~\FINALREF{tab1}, respectively.

 We iteratively calculate (i) $\delta \X$ directly from (\FINALREF{eqsolve}) for each $k$ in terms of ${\bf M}$, 
 (ii) subsequently calculate moments from $\delta \X$ by numerical integration. Importantly, 
 the fix point of the iteration (i)-(ii)-(i)-.. is unique for each $k$, i.e., does not depend on the initial values for moments $A$--$Z$,
 as long as we choose real-valued ones which are consistent with (\FINALREF{eqsolve}), as we prove in the next paragraph.
 In addition, two other computational strategies were implemented: First, we used continuation of functions $A$--$Z$ from their values at $k=0$ to solve (\FINALREF{eqsolve}) with an incremental increase of $k$, where the solution at $k$ was used as the initial guess for $k+dk$. Second, we used also a continuation ``backwards" in which the solution at some $k$ (obtained by convergent iterations with a random initial condition) was used as the initial guess for a solution at $k-dk$. Both these strategies returned the same values of functions $A$--$Z$ as computed by iterations from arbitrary initial condition.
 The solution $\delta \X$ allows to 
 calculate the whole distribution function $f$ via (\FINALREF{expanded}) as illustrated by Fig.~\FINALREF{2007jan18_Fig2}.
 For resulting moments for a wide range of $k$-values see Fig.~\FINALREF{2007jan18_Fig3}.
 
 Finally, we need to clarify the origin of row 5 in Tab.~\FINALREF{tab1}  
 (which is directly illustrated by Fig.~\FINALREF{2007jan18_Fig2}) and its implication on the structure 
 of ${\bf M}$ (\FINALREF{checkerboard}) whose 
 entries are -- a priori -- complex-valued functions
 to be calculated with complex-valued $\delta \X$. 
 	%
 We wish to make use of the fact that all integrals over $z$ vanish for odd integrands. 
 To this end we introduce abbreviations $\plus$ ($\minus$) for a real-valued quantity which is even (odd)
 with respect to the transformation $z\rightarrow -z$. One notices $\X^0=(\plus,\minus,\plus,\plus)$,
 and we recall that $A$--$Z$ are integrals over either even or odd functions in $z$, times a
 component of $\delta \X$ (see Tab.~\FINALREF{tab1}).
 	%
 Let us prove the consistency of the specified symmetry of {\bf M} and the invariance condition: 
 Start by assuming $A$--$Z$ to be real-valued functions. 
 Then $M_{\mu\nu}=\plus$ if $\mu+\nu$ is even, and $M_{\mu\nu}=i\plus$ otherwise. 
 This implies
 $\delta X_1=\plus+i\minus$, $\delta X_2=\minus+i\plus$, $\delta X_3=\plus+i\minus$, and $\delta X_4=\plus+i\minus$,
 i.e., different symmetry properties for real and imaginary parts. 
 With these ``symmetry'' expressions for $\X^0$, $\delta \X$, and ${\bf M}$ at hand, we can insert into 
 the right hand side of the equation,
 $\delta \X = (\X^0+\delta \X)\cdot({\bf M}+i\minus{\bf I})$, which is identical with the invariance equation (\FINALREF{eqsolve}). 
 There are only two cases to consider, because ${\bf M}$ has a checkerboard structure, i.e.,  
 only two types of columns:
 Columns $\mu=1$ and $\mu=3$:
 $\delta X_\mu  
 = \plus + i \minus$ because $M_{1-3,4}=0$;
 Columns $\mu\in\{2,4\}$: $\delta X_\mu 
 = \plus + i \minus$ if $M_{\mu,1-3}=0$ (which is the case for column $4$) 
 and $\minus+i\plus$ if $M_{\mu,4}=0$ (which is the case for column $2$). 
 
          \INLINEFIGURE{
          \begin{figure}[tbh]
          \begin{center}
          \includegraphics[width=8.5cm, angle=0]{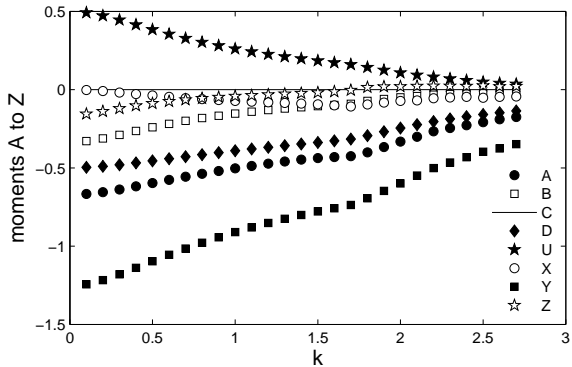}
          \caption{{\footnotesize Moments $A$--$Z$ vs. wave number $k$ obtained with the solution of (\ref{eqsolve}). 
}}
          \label{2007jan18_Fig3}
          \end{center}
          \end{figure}
          }

 We have thus shown that both sides of the invariance equation (\FINALREF{eqsolve}) have equal symmetry properties, 
 and that $\delta \X$ with the specified symmetries is consistent with real-valued 
 moments $A$--$Z$. The proof implies, that any iteratively
 obtained solution, if it exists, starting with arbitrary real-valued moments $A$--$Z$ in (\FINALREF{eqsolve})
 to evaluate $\delta \X$ must converge to real-valued solution $A$--$Z$. Since the solution is smoothly varying
 with $k$, and since $A$--$Z$ at $k=0$ are known and are real-valued, the moments must be 
 real-valued over the whole $k$-space. 
 
 With the result for the functions $A$--$Z$ at hand, the extended hydrodynamic equations are closed.
 Let us briefly discuss the pertinent properties of this system. 
 First, the generalized transport coefficients are given by the nontrivial eigen-values of 
 $-k^{-2}\Re({\bf M})$: $\lambda_2=-A$ (elongation viscosity), $\lambda_3=-\frac{2}{3} Y$ (thermal diffusivity), 
 and $\lambda_4=- D$ (shear viscosity). 
 All these generalized transport coefficients are non-negative (see Fig.\ \FINALREF{2007jan18_Fig3}).
 Second, computing the eigen-values of matrix ${\bf M}$ we obtain the dispersion relation $\omega(k)$ 
 of the corresponding hydrodynamic modes already presented in Fig.\ \FINALREF{2007jan18_Fig1}.
 Third, a suitable transform of the hydrodynamic fields, $\x'={\bf T}\cdot\x$, where ${\bf T}$ is 
 a real-valued matrix, can be established such that the
 transformed hydrodynamic equations, $\partial_t\x'={\bf M}'\cdot{\bf x}'$, with ${\bf M}'={\bf T}\cdot{\bf M}\cdot{\bf T}^{-1}$ 
 is manifestly hyperbolic and stable; 
 $\Im({\bf M}')$ is symmetric, $\Re({\bf M}')$ is symmetric and non-positive semi-definite. 
 The corresponding transformation matrix ${\bf T}$ can be easily read off the results obtained in \cite{matteo1,matteo2} 
 for Grad's systems since the structure of the matrix ${\bf M}$  (\FINALREF{checkerboard}) is identical to the one studied in 
 \cite{matteo1,matteo2}.  We have explicitly verified that matrix ${\bf T}$ (equations (21)--(23) in Ref.\ \cite{matteo1} and (13) in Ref.\ \cite{matteo2}) with the functions $A$--$Z$ derived herein is real-valued and thus render the transformed hydrodynamic equations manifestly hyperbolic and stable.
 We note that this result -- hyperbolicity of exact hydrodynamic equations -- strongly supports a recent suggestion by Bobylev to consider a hyperbolic regularization of the Burnett approximation \cite{Bobylev06}. 
 Similarly, using the hyperbolicity, an $H$-theorem is elementary proven as in \cite{matteo2,Bobylev06}. 
 Finally, using the accurate data for functions $A$--$Z$, 
 we can write analytic approximations for the hydrodynamic 
 equations (\FINALREF{dtx}) in such a way that hyperbolicity and stability is not destroyed in such an approximation (see \cite{matteo1}).

 In conclusion, we derived exact hydrodynamic equations from the linearized Boltzmann-BGK equation.
 The main novelty is the numerical non-perturbative procedure to solve the invariance equation. 
 In turn, the highly efficient numerical approach is made possible by choosing a convenient coordinate system and establishing symmetries of the invariance equation. The invariant manifold in the space of distribution functions is thereby completely characterized, that is, not only equations of hydrodynamics are obtained but also the corresponding distribution function 
 is made available. The pertinent data can be used, in particular, as a much needed benchmark for computation-oriented kinetic theories such as lattice Boltzmann models, as well as for constructing novel models using quadratures in the velocity space \cite{ShanHe98,Chikatamarla06}. Finally, we have established a novel non-perturbative computational approach to finding invariant manifolds of kinetic equations. 
 	 
 The present approach can be extended to the Boltzmann equation with other collision terms.
 The above derivation of hydrodynamics is done under the standard assumption of local equilibrium \cite{chapman},
 however the assumption itself is open to further  
 study \cite{coarsergrainGRAD} [We thank H.C. \"Ottinger for this important remark].
 I.V.K. acknowledges support of CCEM-CH.

 \end{document}